%% file: NeffPaper.tex
\documentclass[traditabstract,10pt]{aa}
\usepackage[T1]{fontenc}
\usepackage{footnote}
\makesavenoteenv{tabular}
\makesavenoteenv{table}
\usepackage[english]{babel}
\usepackage{amsmath, amssymb, amsfonts, latexsym}
\usepackage{epstopdf}
\usepackage{ifthen}
\usepackage[usenames,dvipsnames,table]{xcolor}
\usepackage{fixltx2e}
\usepackage{placeins}
\usepackage{graphicx}
\graphicspath{}
\usepackage{float}
\usepackage[nonamebreak]{natbib}
\usepackage[]{hyperref} 
\def\planck{\textsc{Planck}}

\input{macros}

\begin{document}
\title{Comparison of results on \Neff\ from various \planck\ likelihoods}
\author{S.~Henrot-Versill\'e\thanks{Corresponding author: \href{mailto:versille@lal.in2p3.fr}{versille@lal.in2p3.fr}} \and F. Couchot \and X. Garrido \and H. Imada \and T. Louis \and M. Tristram \and S. Vanneste}
\institute{LAL, Univ. Paris-Sud, CNRS/IN2P3, Universit\'e Paris-Saclay, Orsay, France\label{inst1}
}

\authorrunning{S. Henrot-Versill\'e et al.}

\abstract{
In this paper, we study the estimation 
of the effective number of relativistic species 
from a combination of \CMB\ and \BAO\ data. We vary
different
ingredients of the analysis: 
the \planck\ high-$\ell$ likelihoods,
the Boltzmann solvers,
and the statistical approaches.
The variation of the inferred values 
gives an indication of 
an additional 
systematic uncertainty, which is
of the same order of magnitude as the error derived from each individual likelihood. 
We show that this systematic is essentially associated to the assumptions made in the high-$\ell$ likelihoods 
implementations, in particular for the foreground residuals modellings.
We also compare a subset of likelihoods using only the $\rm{TE}$ power spectra,
expected to be less sensitive to foreground residuals. 
}

\date{\today}

\maketitle 
\section{Introduction}

The expansion rate in the early Universe depends on the energy density of relativistic particles, 
which is parameterised by \Neff, the effective number of relativistic species. 
According to the Standard Model (SM) of particle physics, 
\Neff\ would only receive contributions from the three neutrino species. 
Due to residual interactions, as the neutrinos were not completely decoupled during the 
electron-positron annihilation, \Neff\ is expected to be equal to $3.045$ \citep[]{2016JCAP...07..051D}.

Any deviation from the SM value can be attributed to extra relativistic radiation in the early Universe. 
This can be, for example, massless sterile neutrino species \citep[]{Hamann:2010bk},
axions \citep[]{Melchiorri:2007cd,Hannestad:2010yi}, decay of
non-relativistic matter \citep[]{Fischler:2010xz}, gravitational waves \citep[]{Smith:2006nka,Henrot-Versille:2014jua}, extra
dimensions \citep[]{Binetruy:1999hy,Shiromizu:1999wj,Flambaum:2005it},
early dark energy \citep[]{Calabrese:2011hg}, asymmetric dark matter \citep[]{Blennow:2012de}, or leptonic
asymmetry \citep[]{Caramete:2013bua}. 
Measuring accurately \Neff\ is therefore of particular interest not only to
constrain neutrino physics but also any other process that changes the expansion history.

Any variation of the expansion rate of the Universe affects the \CMB\ power spectra by changing the relative scales
of the Silk  damping relative to the sound horizon \citep[see for instance][]{Abazajian:2013oma}.
Therefore, the current best constraint on \Neff\ comes from the accurate measurements of the temperature 
and polarisation anisotropies performed by \planck.

In this paper we discuss in detail the estimation of \Neff\ from 
\CMB\ data and  quantify the dependence of the results on the choices made in the analysis. 
We investigate different possible sources of systematic errors. 
We first compare the results
obtained using two Boltzmann codes: \CAMB\ \citep[][]{camb} and \CLASS\ \citep[][]{2011JCAP...07..034B}. We then use
three different \planck\ high-$\ell$ likelihoods. 
We also discuss the statistical analysis, comparing the frequentist and Bayesian approaches, 
to pin-point any remaining volume effects. We show that varying the above listed ingredients 
lead to a non-negligible spread of the mean \Neff\ values.

The paper is organised as follows.
In Sect.~\ref{methodo}, we introduce the datasets, the \planck\ likelihoods, the Boltzmann codes and the statistical analysis.
In Sect.~\ref{ALLsec}, we quantify the effect of possible sources of systematic error on \Neff\ using the combination
of temperature and polarization CMB data ($\rm{TT+TE+EE}$) together with Baryon Acoustic Oscillation (\BAO) data.
In Sect.~\ref{consistencysec}, we compare the results obtained with the CMB $\rm{TT}$
and  $\rm{TE}$ power spectra. The conclusions are given in Sect.~\ref{conclusec}.

\section{Phenomenology and Methodology}
\label{methodo}
\subsection{Introduction}
\Neff\ stands for the effective number of relativistic degrees of freedom. It relates
the radiation ($ \Omega_{\rm rad}$) and the photon ($\Omega_{\gamma}$) energy densities relative to the critical density
through:
\begin{equation} \label{eq:Neff}
  \Omega_{\rm rad}=\left(1+\frac{7}{8}\Neff\left(\frac{4}{11}\right)^{4/3} \right) \Omega_{\gamma} \ .
\end{equation}

Under the assumption that only photons and standard light neutrinos 
contribute
to the radiation energy density, $\Neff$ is equal to the effective number
of neutrinos: $\Neff\simeq 3.045$.
This value has been derived from the number of neutrinos
constrained by the measurement of the decay width of the
Z boson \citep[]{Beringer:1900zz}, and takes
into account residual interactions during 
the electron-positron annihilation.

\subsection{Data sets and likelihoods}
\label{likdescri}
The datasets and likelihoods that have been used in this paper
are summarized together with their corresponding acronyms in \mytable~\ref{tab:acro}. 
Several high-$\ell$ (respectively low-$\ell$) likelihoods have been derived from 
the \planck\ 2015 data \citep{planck2014-a13,Couchot:2016vaq,planck2014-a25},
they are further described in Sect.~\ref{sec:highell}  (respectively Sect.~\ref{sec:lowell}).
The Baryon Acoustic Oscillation data are also discussed in Sect.~\ref{sec:bao}.
 
\begin{table*}[t]
\centering          
\begin{tabular}{l l } 
\hline\hline       
Acronym & Description  \\
\hline                    
$\hbox{\hlpXXps}$ &   high-$\ell$  \hillipopps\  \planck\ likelihood \\ 
$\hbox{\CamSpec} $& Cambridge high-$\ell$  \planck\ likelihood  \\ 
$\hbox{\Plik} $& public high-$\ell$  \planck\ likelihood  \\ 
\hline
$\hbox{\hlpXX }$&  high-$\ell$ \hillipopps\  \planck\  likelihood (one point source amplitude per cross-spectrum)\\ 
$\hbox{\hlpXXps(\Plik-like) }$&  high-$\ell$ \hillipopps\  \planck\  likelihood (see Section \ref{results})\\ 
\hline
$\hbox{TT }$& refers to the temperature power spectra  \\ 
$\hbox{TE }$& refers to the temperature and E modes cross-spectra \\
$\hbox{EE }$& refers to the  E modes power spectra\\
$\hbox{ALL}$& refers to the combination of temperature and polarisation CMB data (incl. $\rm{TT}$, $\rm{TE}$, and $\rm{EE}$)\\
\hline
$\hbox{Comm} $& \Commander\  low-$\ell$ temperature \planck\ public likelihood  \\
$\hbox{\lowTEB} $& pixel-based temperature and polarisation low-$\ell$ \planck\ public likelihood \\
\hline
$\hbox{\BAO} $& Baryon Acoustic Oscillation data (cf. Sect.~\ref{sec:bao}) \\
\hline            
$\hbox{PLA}$ & \href{http://www.cosmos.esa.int/web/planck/pla}{Planck Legacy Archive}  \\
\hline
\end{tabular}
\caption{Summary of keywords, data and likelihoods together with their corresponding acronyms used in this paper.  
\Plik, \Commander, \lowTEB\ 
are the public likelihoods delivered by the \planck\ consortium. See text for detail and references.\label{tab:acro} }  
\end{table*}

\subsubsection{low-$\ell$ likelihoods}
\label{sec:lowell}

At low multipoles ($\ell<50$), the \planck\  public likelihood is \lowTEB, based on \planck\
Low Frequency Instrument (LFI) maps
at 70GHz for polarization and a component-separated map using all
\planck\ frequencies for temperature \citep[\Commander][]{planck2014-a13}.

In the following, we have also tested a combination of the \lollipop\ likelihood \citep{lollipop}
with \Commander\ in place of \lowTEB, following what has been done for 
the latest \planck\ results on the reionisation optical depth \citep{planck2014-a25}.

\subsubsection{high-$\ell$ likelihoods}
\label{sec:highell}

At high multipoles ($\ell>50$), different likelihoods were developed within the \planck\ collaboration: 
\plik\ \citep[][]{planck2014-a13} being the one delivered to the community. 
Their implementations are further detailed in this Section. 
Since there is no valuable reason to favor an implementation or another,
we use them in the following to assess the impact of the various ingredients entering
their derivation on the \Neff\ inferred value. We consider 
\plik, \CamSpec\ \citep[][]{planck2014-a13}, and \hillipopps.

All those likelihoods are based on pseudo-$C_\ell$ cross-spectra between \planck\ High Frequency Instrument 
(HFI)
half-mission 
maps at 100, 143 and 217~GHz \citep[for more details, see][]{planck2014-a13}. The main differences are listed below:
\begin{itemize}
\item {\bf Data} \hillipopps\ makes use of all 15 cross-spectra from the 6 half-mission maps whereas \plik\ and \camspec\ remove the $100\times143$ and $100\times217$ correlations together with two of the four $143\times217$ cross-spectra (for temperature data only). To avoid residual contamination from dust emission, \hillipopps\ and \camspec\ do not use the multipoles below 500 for the $143\times217$ and $217\times217$ cross-spectra.
\item {\bf Masks} The Galactic masks used in temperature are very similar. Still, \hillipopps\ relies on a more refined procedure for the point source masks that preserves Galactic compact structures and ensures the completeness level at each frequency, but with a higher detection threshold (thus leaving more extra-Galactic diffuse sources residuals). In polarization, \camspec\ uses a cut in polarization amplitude ($P=\sqrt{Q^2+U^2}$) to define diffuse Galactic polarization masks whereas \hillipopps\ and \plik\ use the same masks as in temperature.
\item {\bf Covariance matrix } The approximations used to calculate the covariance matrix which encompasses the $\ell$-by-$\ell$ correlations between all the cross-power spectra are slightly different. \plik\ and \camspec\ assume a model for signal (from cosmological and astrophysical origin) and noise (with slight differences in the methods used to estimate noise). In \hillipopps, it is estimated semi-analytically with \xpol\ 
\citep[a polarized version of the power spectrum estimator described in][]{Xspect} using a smoothed version of the estimated spectra \citep{Couchot:2016vaq}.
\item {\bf Galactic dust template} \hillipopps\ uses templates for the Galactic dust emission derived from \planck\ measurements both for the shape of the power spectra \citep{planck2014-XXX} and for the 
Spectral Energy Distribution \citep{planck2014-XXII}, rescaled by one amplitude for each polarisation mode ($\rm{TT}$, $\rm{EE}$ and $\rm{TE}$). In contrast, due to Galactic cirrus residuals that are included in their point source masks, \plik\ and \camspec\ have to rely on an empirical fit of the spectrum mask difference at 545~GHz and fit one amplitude for each of the cross-frequency spectra with priors on the amplitude based on a power-law (with slightly different spectral index: $-2.63$ for \plik\ and $-2.7$ for \camspec). In polarization, \camspec\ compresses all the frequency combinations of $\rm{TE}$ and $\rm{EE}$ spectra into single $\rm{TE}$ and $\rm{EE}$ spectra (weighted by the inverse of the diagonals of the appropriate covariance matrices), after foreground cleaning using the 353\,GHz maps. As a consequence, \camspec\ has no nuisance parameters describing polarized Galactic foregrounds.
\item {\bf SZ template} The template spectra for thermal Sunyaev-Zeldovich (SZ) effect residuals is based on a model for \plik\ and \camspec; whereas it comes directly from \planck\ measurements in the case of \hillipopps.
\item {\bf Point sources template} \hillipopps\ 
includes a 2-components point source model (including infrared dusty galaxies and extragalactic radio sources) with one amplitude for each component and 
a fixed SED whereas all the other likelihoods fit one point source amplitude for each cross-frequency. We also consider a version which fits one point source amplitude per cross-spectrum (as what is done in \Plik), labelled \hillipop.
\end{itemize}

The results obtained with those high-$\ell$ likelihoods have been
compared in  \citet[]{planck2014-a13} when combined with a prior on 
the optical depth to reionization ($\taureio$).
It was shown that the \lambdaCDM\  parameters derived
from temperature data were very compatible. Still, as described in \citet[]{couchot:2015},
when combining them with \lowTEB,
a disagreement was observed, especially on $\taureio$ and $\As$ 
(the initial super-horizon amplitude of
curvature perturbations at k = 0.05 Mpc$^{-1}$).
This was shown to be related to a discrepancy on the 
$\Alens$ fitted value:$\Alens$ is a phenomenological
parameter that was first introduced in \citet[]{Calabrese:2008rt}
to cross-check the consistency
of the data with the $\lambdacdm$ model.
Further studies on the impact of
those differences
on the measurement of the sum of the neutrino mass were
also performed in \citet[]{Couchot:2017pvz}. 

In the following we investigate the systematic effects hidden
in the assumptions made for the derivation of those likelihoods.
As the use of a single likelihood does not ensure the full propagation
of errors, we base our analysis on a comparison of the results inferred
from each of them and
estimate the order of magnitude of the related errors.

\subsubsection{Baryon Acoustic Oscillation (BAO) data}
\label{sec:bao}

Informations on the late-time evolution of the Universe geometry
are also included. In this work, 
we use 
the acoustic-scale distance ratio $D_{\hbox{\tiny{V}}}(z)/r_{\mathrm{drag}}$
measurements 
from the 6dF Galaxy Survey at $z=0.1$ \citep{Beutler:2014yhv}. 

$D_{\hbox{\tiny{V}}}(z)$ is a combination of the comoving angular diameter
distance $D_{\hbox{\tiny{M}}}(z)$ and Hubble parameter $H(z)$ according to:
\begin{equation}
D_{\hbox{\tiny{V}}}(z) = \left[ D^2_{\hbox{\tiny{M}}}(z)  \frac{cz}{H(z)}\right]^{1/3}
\end{equation}
and $r_{\mathrm{drag}}$ 
is the comoving sound horizon at the end of
the baryonic-drag epoch.
At higher redshift, we have also included the 
BOSS DR12 \BAO\ measurements \citep{Alam:2016hwk}. They consist in constraints on $ (D_M(z), H(z), f(z)\sigma_8(z)) $ 
in three redshift bins, which encompass both BOSS-LowZ  and   BOSS-CMASS DR11 results. 
$\sigma_8(z)$ gives the normalization of the linear theory matter power spectrum 
at redhift $z$ on 8$h^{-1}$ Mpc scales. $f(z)$ is
the derivative of the logarithmic growth rate of the linear fluctuation amplitude with respect to 
the logarithm of the expansion factor.
The combination of those measurements is  labelled  \BAO\ in the following.
We note that this is an update of the \BAO\ data with respect
to those used in \citet{planck2014-a15}.

\subsection{Statistics and Boltzmann codes}
\label{neutr}

We use the \CAMEL\ software\footnote{\url{camel.in2p3.fr}}\citep[]{Henrot-Versille:2016htt} tuned to a high precision setting
to perform the statistical analysis.
It allows us to compare both the frequentist (profile likelihoods) and the Bayesian approaches.
\CAMEL\ includes a MCMC algorithm based on the Adaptative Metropolis method \citep[]{haario2001}. It
also encapsulates the 
\CLASS\ Boltzmann solver \citep{2011JCAP...07..034B}. 
The \CLASS\ and \CAMB\ softwares have been 
extensively compared \citep[]{class-camb}, and lead
to very close predictions in terms of CMB spectra. 
Still, the public \planck\
results are derived using \CAMB: their comparison
with the ones derived with \CAMEL\ allows us to cross-check the compatibility of the theoretical predictions
while fitting for \Neff. 

For both setups, we are using the 
model of  \citet{2012ApJ...761..152T} extended to massive neutrinos
as described in  \citet[]{2012MNRAS.420.2551B} to include 
non-linear
effects on the matter power spectrum evolution.
We have used the big-bang nucleosynthesis
(BBN) predictions
calculated with the PArthENoPE code \citep[][]{Consiglio:2017pot} updated to the 
latest observational data on nuclear rates 
and assuming a neutron lifetime of 880.2 s,
identical to the standard assumptions made in \citet[][]{planck2014-a15}.

For \lambdaCDM, 
we assume that all the neutrino mass (\mnu=0.06~\eV) is carried out by only one heavy neutrino. 
Considering today's knowledge on the neutrino sector \citep[][]{PhysRevD.98.030001},
we do not, yet, have access to a measurement of the individual masses.
Two mass hierarchy scenarios are therefore considered in the literature:
the normal hierarchy
with $m_1 < m_2 \ll m3$ and the inverted hierarchy
with $m_3 \ll m_1 < m_2$, where $m_i$ (i = 1; 2; 3)
denotes the neutrinos mass eigenstates. 
In this paper, we have also performed fits with three neutrinos with a mass splitting scheme derived 
from the normal hierarchy (keeping \mnu=0.06~\eV)
and we did obtain identical results.

\begin{figure}[!ht]
\centering
\includegraphics[width=8.5cm]{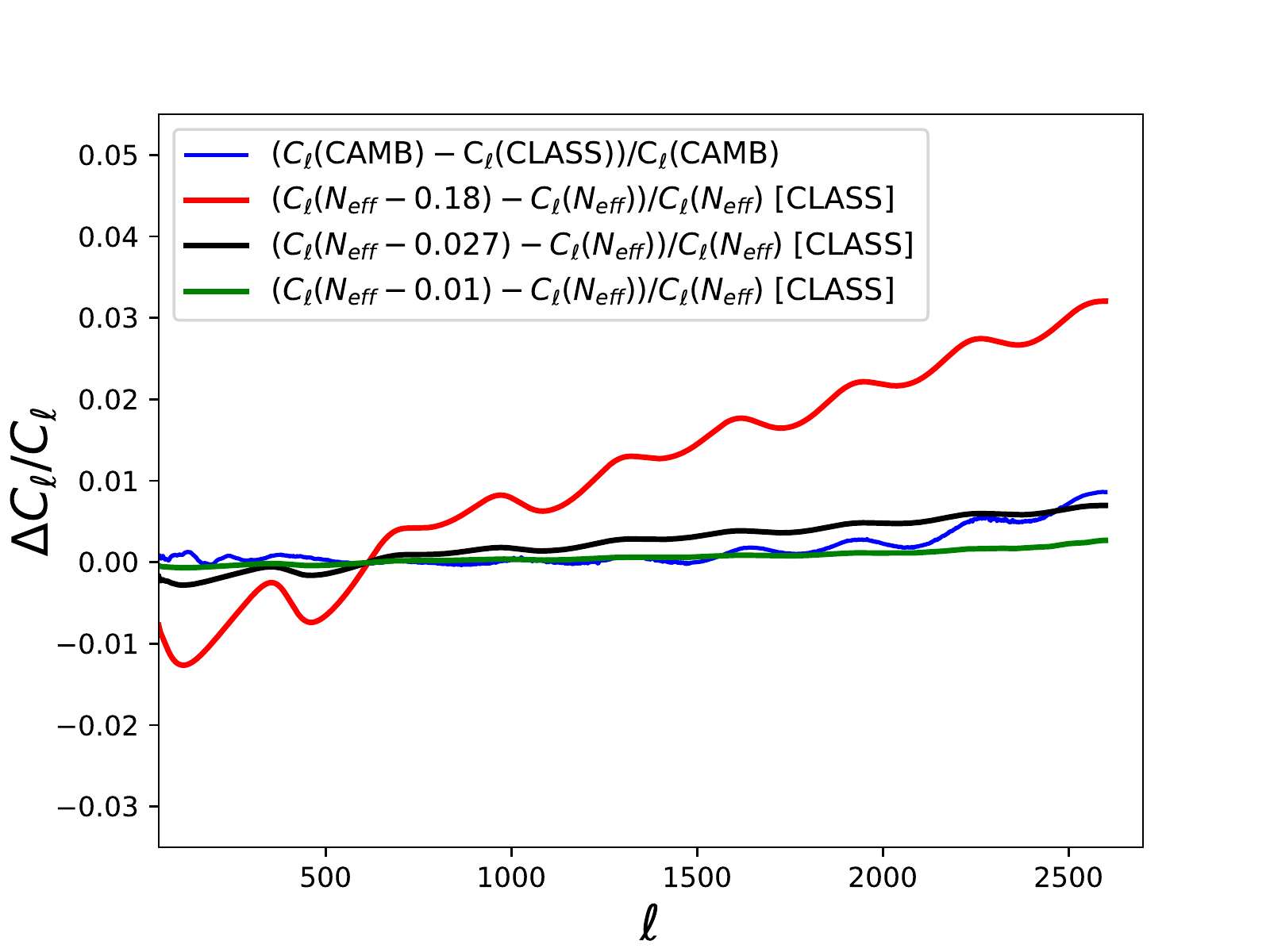}
\caption{Relative variations of the predicted temperature spectra between \CLASS\ and \CAMB\ (blue).
We also compare the spectra when we shift \Neff\  toward negative values for
$\Delta\Neff=-0.18$  (red), $\Delta\Neff=-0.027$ (black) 
and $\Delta\Neff=-0.01$ (green) 
with \CLASS\ ($\theta$ is fixed, $H_0$ is
therefore recalculated).}
\label{fig:compare_spec}
\end{figure}

We illustrate in \myfigure~\ref{fig:compare_spec} 
the relative variations of the temperature spectra ($\Delta C_\ell/C_\ell$)
between \CAMB\ and \CLASS. We show the impact of a negative shift of the \Neff\
value in three cases: $\Delta\Neff=-0.18$ corresponding to the $1\sigma$ error reported by \planck,
$\Delta\Neff=-0.027$ which is the forecasted uncertainty for the next
generation 'Stage-4' ground-based CMB experiment, 
CMB-S4 \citep[][]{Abazajian:2016yjj},
and $\Delta\Neff=-0.01$ which is close to the \CAMB/\CLASS\ difference.
The non-linear effects have been deliberately
neglected to produce this Figure.

\section{$\rm{TT}$+$\rm{TE}$+$\rm{EE}$+BAO results}
\label{ALLsec}
In this section, we discuss the results obtained with the combination of \planck\ $\hbox{TT+TE+EE}$  (so-called $\hbox{ALL}$) likelihoods 
together with \BAO\ data. 
They are given in \mytable~\ref{tab:ALL}, and
are classified according to various kind of systematic errors. Each of them is further discussed in a dedicated subsection below:
we first assess the impact of the choice of the Boltzmann solver, then we discuss the impact
of the choice of the high$-\ell$ likelihood. Finally we compare the results using different statistical analysis.
 
All the values which are tagged with a \offismall\ are extracted from the 
PLA.

\begin{table*}
\centering          
\begin{tabular}{l l l c } 
\hline\hline       
& \hbox{\planck}\ $\cal{L}$ & Config & \Neff \\
& \hbox{+lowTEB+BAO} &  &  \\
\hline
1 & $\hbox{\PlikALL\offi}$ &MCMC/\CAMB &  $3.04\pm0.18$  \\
\hline
& {\bf Boltzmann code and sampler systematics }  \\
\hline
2 & $\hbox{\PlikALL}$ &MCMC/\CLASS &  $3.03^{-0.17}_{+0.17}$   \\ 
\hline
& {\bf Likelihood systematics } \\
\hline
3 & $\hbox{\CamSpecALL\offi}$ & MCMC/\CAMB & $2.89\pm0.19$ \\
4 & $\hbox{\hlpALL}$ &MCMC/\CLASS &   $2.92^{-0.15}_{+0.15}$   \\ 
5 & $\hbox{\hlpALLps}$ &MCMC/\CLASS &    $2.86^{-0.14}_{+0.15}$  \\ 
\hline
& {\bf Statistical analysis systematics } \\
\hline
6 & $\hbox{\PlikALL}$ &Profile/\CLASS &  $3.00_{-0.20}^{+0.19}$  \\ 
7 & $\hbox{\hlpALL}$ &Profile/\CLASS &  $2.87_{-0.14}^{+0.15}$    \\ 
8 & $\hbox{\hlpALLps}$ &Profile/\CLASS &  $2.85\pm 0.14$    \\ 
\hline
9 & $\hbox{\hlpALLps (\plik-like)}$ &Profile/\CLASS & $2.90_{-0.16}^{+0.17}$ \\ 
\hline
\end{tabular}
\caption{Results on \Neff\ obtained when combining  \PlikALL, \hlpALL\ and \hlpALLps\ 
with \BAO (errors are given at 68$\%$CL).
\lowTEB\ has been used at low-$\ell$.  \label{tab:ALL}}
\end{table*}

\subsection{Boltzmann code and sampler effects}

In this subsection, we study the impact of the choice of the 
Boltzmann solver that is used to infer cosmological parameters.
Within our setup we
cannot disentangle the impact of the Boltzmann code from the one of the sampler used
for the MCMC mutiparameter space exploration, 
as a consequence the estimation given in this section combine both effects. 

The comparison of the results using \Plik\ are
given in \mytable~\ref{tab:ALL} (line 1 and 2): the use of \CLASS\ combined with the \CAMEL\ 
MCMC sampler tends to induce slightly smaller error bars on \Neff\ as well as a very small
shift of $0.01$ toward lower values when results are compared with the public \planck\
results. It is further illustrated by the difference between the black (for the public/\CAMB) and the
blue (for this work/\CLASS) marginal distributions on cosmological parameters
shown on \myfigure~\ref{fig:compare_MCMC} (see next section for a full description
of the Figure). This shift is consistent with the difference shown on
\myfigure~\ref{fig:compare_spec} between spectra predicted by both Boltzmann solvers,
and is largely subdominant compared to the statistical uncertainty. 

We also tested the effect of changing the neutrino model.
We have compared the 
results when attributing to each of the three neutrinos a mass derived 
from the Normal Hierarchy scenario expectation and found a $0.01$ shift
of the \Neff\ results. Given the actual precisions on the \CMB\ spectra,
we can therefore safely assume a \lambdaCDM\ model with 
only one massive neutrino 
carrying all the mass.

\subsection{Likelihood comparisons}
\subsubsection{Results}
\label{results}
A possible source of  systematic error to be estimated is the one related to the choice 
of the \planck\ high-$\ell$ likelihood. As discussed in Sect. \ref{likdescri}, 
various
assumptions have been made to build the likelihood. 
The
comparison of the results from each likelihood allows to quantify the 
impact of the underlying assumptions.

A discrepancy between \PlikALL\ and \CamSpecALL\ is already mentioned
in \citet[]{planck2014-a15}, which quotes  
$\Delta\Neff\simeq 0.15$. 
Using the \hlp\ likelihoods, we find differences of the same order
of magnitude as 
quoted in \mytable~\ref{tab:ALL} (line 1 vs. lines 3,4,5). 
This variation can reach a maximum of $\Delta\Neff\simeq 0.17$.

However, as stated in Sect.~\ref{likdescri}, there are more data in the \hillipopps\ likelihoods
than in \plik\ and \camspec. This can affect the interpretation of the shift,
as part of it might be due to statistical fluctuations.
To test this effect, we have derived the results using \hlp\ while removing 
the $100\times143$, $100\times217$ and two of the four $143\times217$ cross-spectra,
and reducing the $\ell$ range (cf. Sect.~\ref{sec:highell}): the result is 
quoted on line 9 and labelled \hillipopps(\plik-like). 
We see that a small part (up to $0.03$) of those $0.17$ may be attributed
to a statistical effect (including the covariance matrix determination). 

\subsubsection{Correlations with other parameters}

In this section we investigate the correlation between \Neff\ and
the cosmological and nuisance parameters, the definition
of the lattest being given in Table \ref{tab:hlp_nuisance}. For \hlpPS, 
the model for the point source residuals is slighly different (see Section \ref{sec:highell}): 
$A_{\hbox{radio}}$ and $A_{\hbox{dusty}}$ are respectively the 
amplitudes of the radio sources and the dusty galaxies \citep[][]{TENous}.
The nuisance parameters for \Plik\ are further defined in 
\citet{planck2013-p08}
and \citet{planck2014-a13}.
The cosmological parameters we infer together with \Neff\ are the sixth
parameters of the base \lambdacdm\ model, as defined in \citet{planck2013-p11}, namely:
\begin{itemize}
\item{} $\Omega_bh^2$: Today's baryon density
\item{} $\Omega_ch^2$: Today's cold dark matter density 
\item{} $H_0$: Current expansion rate in $\hbox{km.s}^{-1}\hbox{.Mpc}^{-1}$
\item{} $\taureio$: Optical depth to reionization
\item{} $n_s$: Scalar spectrum power-law index
\item{} $\ln(10^{10}A_s)$: Log power of the primordial curvature perturbations
\end{itemize}

\begin{figure*}[!ht]
\centering
\includegraphics[scale=.6,angle=-90]{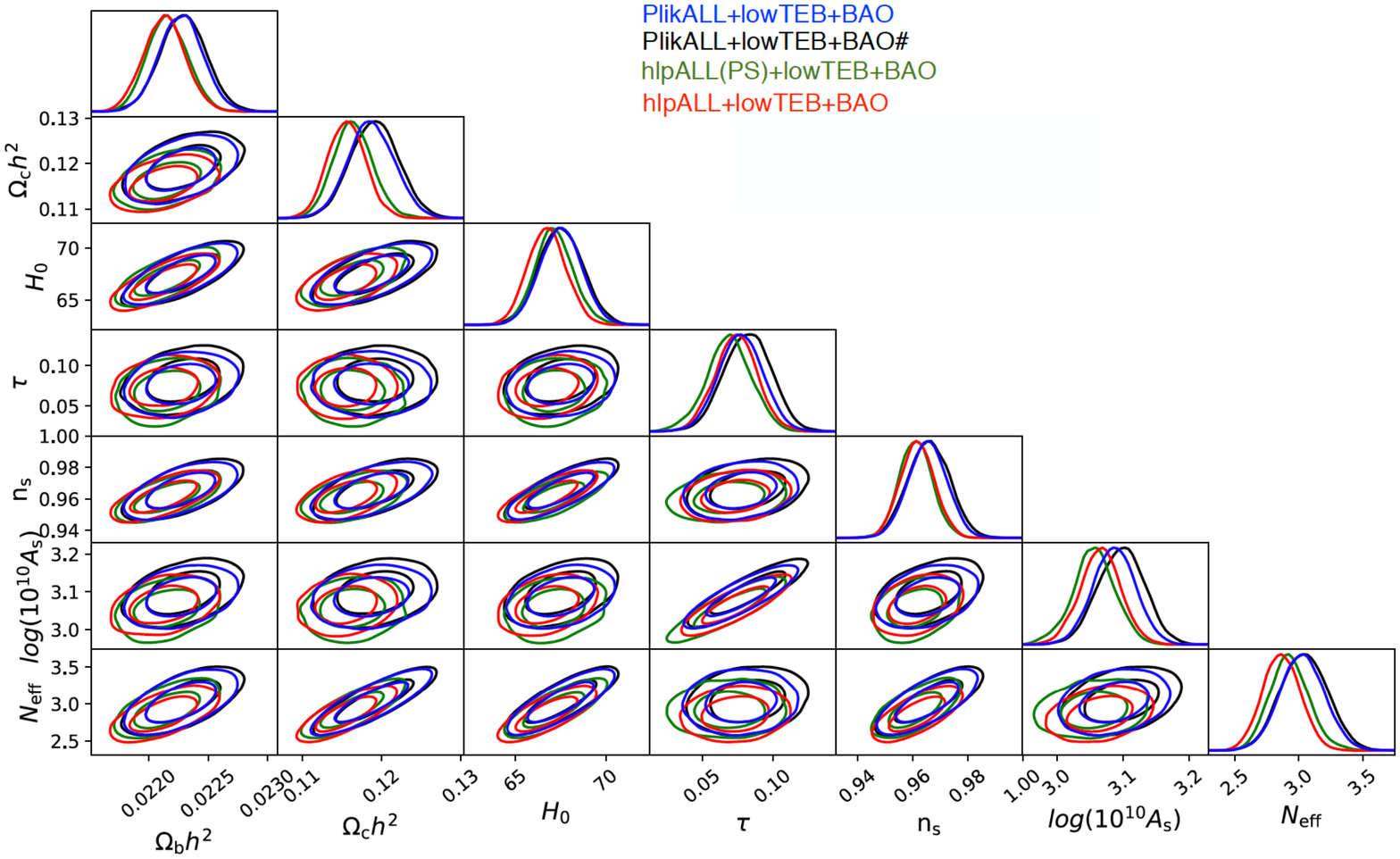}
\caption{Cosmological parameters obtained with a combination of \lowTEB, \BAO and 
high$-\ell$ likelihoods: \hlp, \hillipopps, \plik/\CLASS\ 
and \plik/\CAMB\ (the chains are the ones of the PLA using \CAMB). }
\label{fig:compare_MCMC}
\end{figure*}

We give on \mytable~\ref{tab:param} and \myfigure~\ref{fig:compare_MCMC}
the results of the \CAMEL\ MCMC sampler 
using the \CLASS\ Boltzmann
solver with \PlikALL, \hlpALL\ and \hlpALLps\ (combined with \lowTEB\ 
and \BAO) compared to the \planck\ public chains for \Neff\ plus
the six \lambdacdm\ parameters. Similarly to what is observed on \Neff, 
we find variations of the parameters between likelihoods of the order of one sigma
or less.
The error bars of the \hillipopps\ likelihoods are slightly smaller due to the additional data that
are used (cf. Sect.~\ref{sec:highell}).

\begin{table*}
\centering          
\begin{tabular}{l c c c c } 
\hline\hline       
Param &   \hlpALL[\CLASS] &   \hlpALLps[\CLASS] &   \plikALL[\CLASS] & \plikALL[\CAMB]\offi \\
\hline
$\Omega_bh^2$   & $0.02215^{-0.00017}_{+0.00018}$ & $0.02213^{-0.00018}_{+0.00017}$ & $0.02227^{-0.00018}_{+0.00018}$ & $0.02229^{-0.00019}_{+0.00019}$ \\
$\Omega_ch^2$   & $0.1163^{-0.0023}_{+0.0025}$ & $0.1155^{-0.0023}_{+0.0023}$ & $0.1187^{-0.0029}_{+0.0030}$ & $0.1191^{-0.0031}_{+0.0030}$ \\
$H_0$           & $67.14^{-1.01}_{+1.074}$ & $66.71^{-1.02}_{+1.058}$ & $67.51^{-1.13}_{+1.125}$ & $67.49^{-1.21}_{+1.235}$ \\
$\taureio$      & $0.069^{-0.016}_{+0.016}$ & $0.074^{-0.015}_{+0.014}$ & $0.077^{-0.016}_{+0.016}$ & $0.082^{-0.017}_{+0.017}$ \\
$n_s$           & $0.961^{-0.006}_{+0.006}$ & $0.962^{-0.006}_{+0.006}$ & $0.965^{-0.007}_{+0.007}$ & $0.966^{-0.008}_{+0.008}$ \\
$\ln(10^{10}A_s)$ & $3.06^{-0.03}_{+0.03}$ & $3.07^{-0.03}_{+0.03}$ & $3.09^{-0.03}_{+0.03}$ & $3.10^{-0.04}_{+0.03}$ \\
$\Neff$  & $2.92^{-0.15}_{+0.15}$ & $2.86^{-0.14}_{+0.15}$ & $3.03^{-0.17}_{+0.17}$ & $3.04^{-0.18}_{+0.18}$ \\
\hline
\end{tabular}
\caption{Results on MCMC chains for all cosmological parameters
obtained when combining the \PlikALL, \hlpALL\ and \hlpALLps\  \planck\ 
likelihoods with \lowTEB\ and \BAO\ (errors are given at 68$\%$CL).\label{tab:param}} 
\end{table*}

\begin{figure*}[!ht]
\centering
\includegraphics[width=12.5cm,angle=-90]{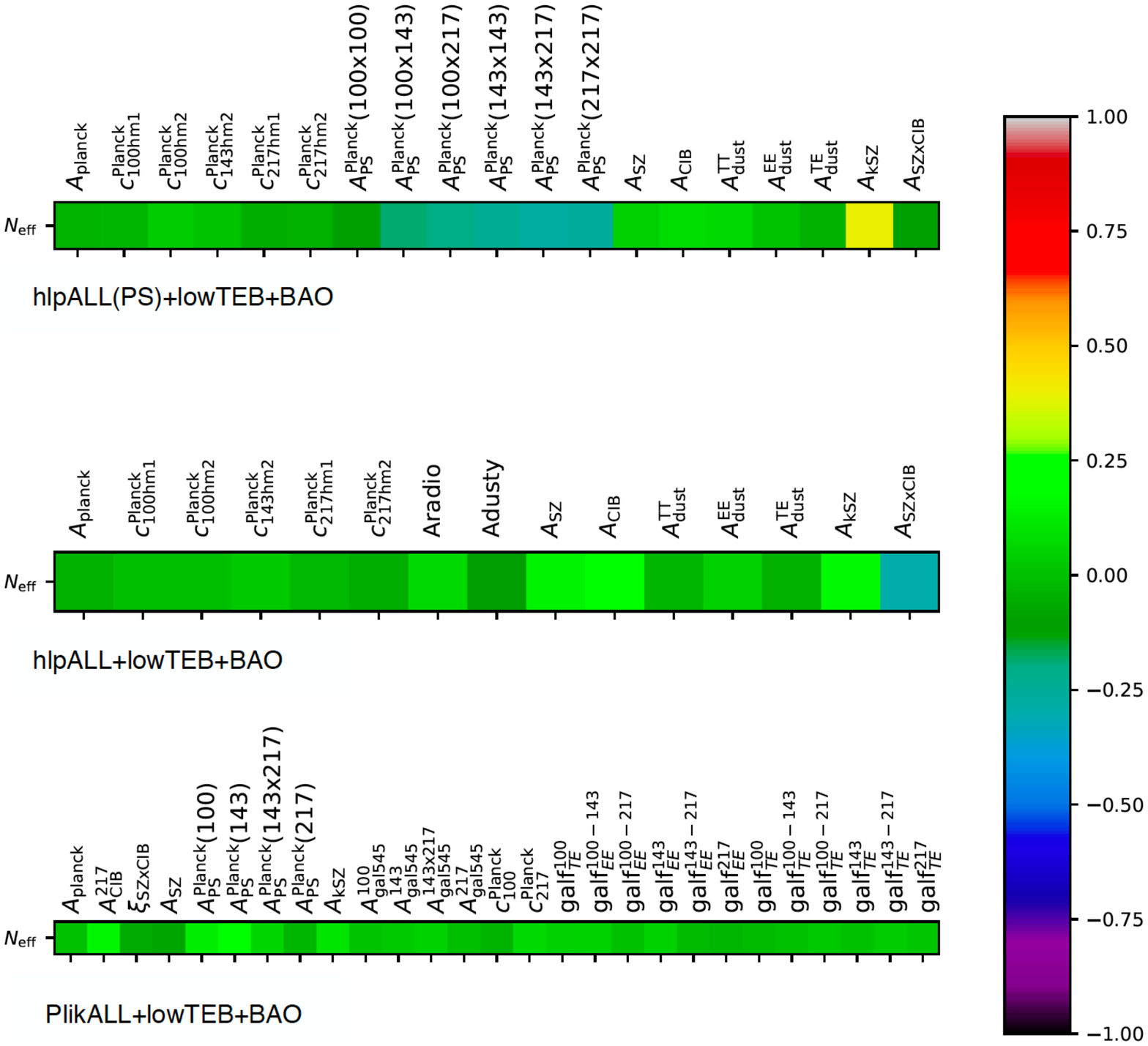}
\caption{Correlation coefficients between \Neff\ and the nuisance parameters
from top to bottom for \hlpALL,
\hlpALLps\ and \plikALL\ when combined with \BAO\ and \lowTEB. }
\label{fig:CoeffCorr}
\end{figure*}

The correlations between \Neff\ and the nuisance parameters are
illustrated on \myfigure~\ref{fig:CoeffCorr} for the three likelihoods (from top to bottom: \hlpALL, \hlpALLps, \plikALL).
For \hlp,
the highest values of the coefficients are obtained for the nuisance parameters 
related to foregrounds which play a role at small scales: namely  the point sources (with a ``PS'' label
in the name of the parameter) and/or the
SZ sector. \Neff\ is anti-correlated to the point source parameters when the related nuisance parameters
are left free to vary (as this is the case of \hillipopps). While, adding information in the point source model
(as done in \hillipopps), this relation is broken.  We also observe a correlation between \Neff\ and $A_{\rm{kSZ}}$
(the amplitude
of the kinetic SZ effect)
and $A_{\rm{SZxCIB}}$ (the amplitude of the correlation between SZ and the Cosmic Infrared Background CIB) 
but those parameters are only very slightly constrained with
\planck\ data. For \plik, the correlation level is lower for all nuisance parameters, 
but the number of parameters is higher.

\subsection{Statistical analysis systematics}
\subsubsection{Results}
\begin{figure}[!ht]
\centering
\includegraphics[width=6.cm,angle=-90]{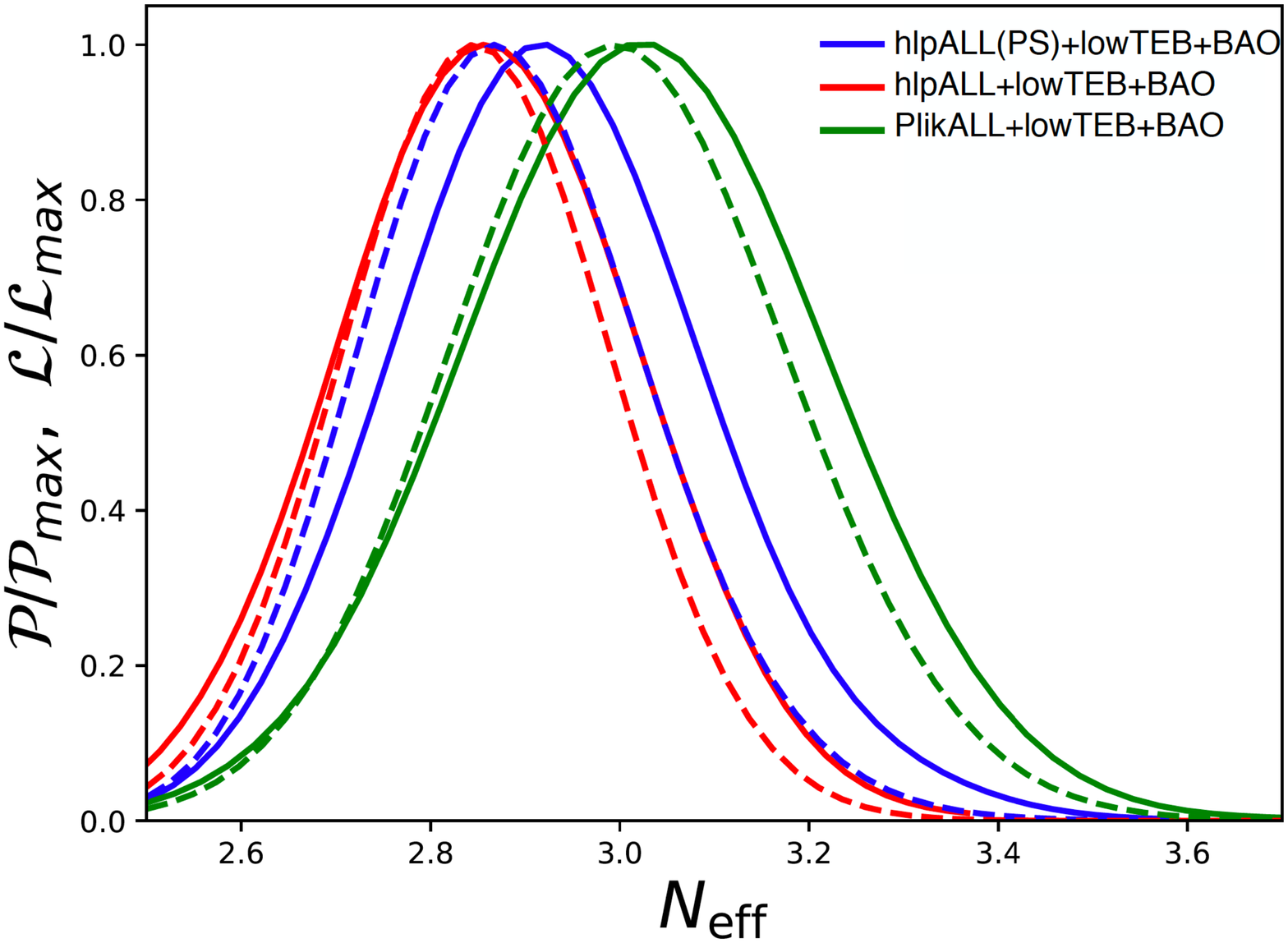}
\caption{Posterior distributions of the MCMC analysis
(plain lines) and Profile likelihood ratio ${\cal{L}/\cal{L}}_{max}$  (dashed lines) for 
\hlpALL\ (blue), \hlpALLps\ (red), and \plikALL\ (green) combined with \lowTEB+\BAO.}
\label{fig:compare_stat}
\end{figure}

In this section, we sudy the impact of the choice of the statistical analysis 
(Bayesian vs. frequentist). The main purpose
of such a comparison is to check for any volume effect that may impact
significantly the results \citep[see for example][]{BayesianPourri}. 
The  \Neff\ estimates for various \planck\ likelihoods
using profile likelihoods are given on lines 6 to 8 of \mytable~\ref{tab:ALL}.
A visual comparison of the results are shown on \myfigure~\ref{fig:compare_stat}, 
where the profile analysis results are transformed in terms of ${\cal{L}/\cal{L}}_{max}$ 
and are superimposed to the MCMC posterior distributions.

The profile analysis results systematically lead to smaller
mean values, keeping the error bars almost similar. 
This effect is also present in the PLA: for example, the
\Neff\ values extracted from the best fit procedure (which is exactly
what is done in a profile analysis) quoted for the \PlikALL+\lowTEB+\BAO\ combination
is equal to $2.996$, a value which is $\simeq 0.04$ smaller than the
maximum of the MCMC posterior distributions. The variation is
specific to each likelihood and not expected to be constant as
it reflects its very shape in the multidimensional parameter space. The higher volume 
effect is observed for \hlp\ and does
not exceed $\Delta\Neff=0.05$. 

\subsubsection{Statistical and Nuisance error contribution}
\label{syst-statsec}

\begin{figure}[!ht]
\centering
\includegraphics[width=6.5cm]{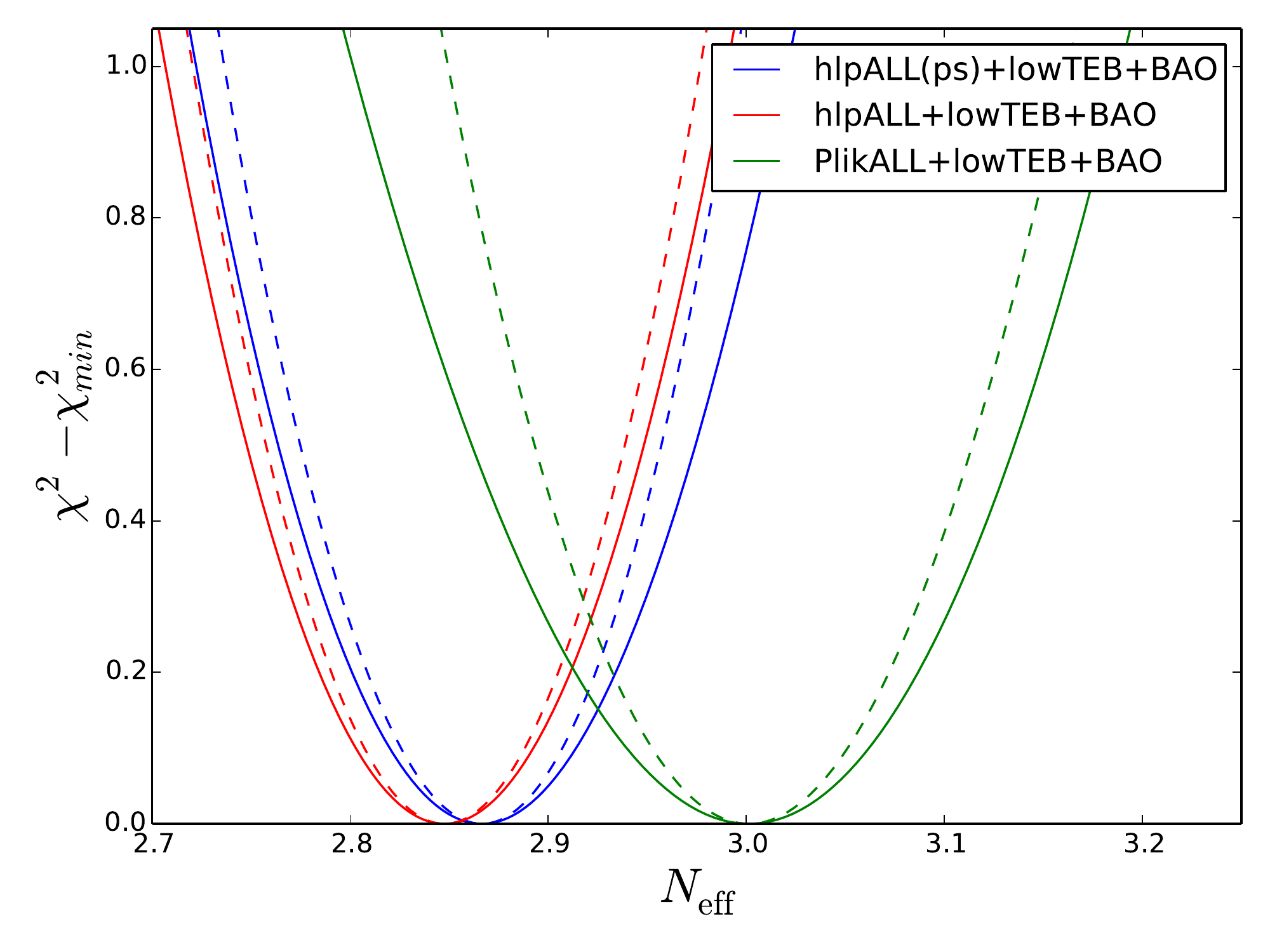}
\caption{Comparison of profile likelihoods obtained for the combination of \hlpALL\ (blue),
\hlpALLps\ (red) and \plikALL\ (green) together with \BAO\ and \lowTEB. The
superimposed dashed profiles have been obtained when fixing the nuisance parameters values
to the ones obtained for the best fit of each likelihood combination.}
\label{fig:Syst-Stat}
\end{figure}

Following the procedure described in  \citet[][]{Aad:2014aba}, we 
have separately estimated the two contributions to the total error: 
the one coming  from statistics and 
the one linked to the foreground and instrumental modelling
(so-called nuisance error).
We first built the usual profiles for each likelihood: they
are shown in solid lines on \myfigure~\ref{fig:Syst-Stat} and the corresponding results
are given in lines 6 to 8 of \mytable~\ref{tab:ALL}. In a second step,
we built another set of profiles, fixing the nuisance parameters to 
the values of the previously obtained best-fit. The errors derived from this second fit (shown in dashed line
on \myfigure~\ref{fig:Syst-Stat})  correspond to the ultimate error
one would obtain if we knew the nuisance parameters perfectly (and they had the values given by the best-fit). 
Finally the nuisance error of each individual likelihood is deduced by quadratically subtracting the 
statistical uncertainty from the total error. The results are given in \mytable~\ref{tab:syst-stat}.

\begin{table}
\centering          
\begin{tabular}{l l l l l } 
\hline\hline       
\hbox{\planck}\ $\cal{L}$ & Mean  & Full  & Stat  & Nuisance  \\
\hbox{lowTEB+BAO} & Value & Error & Error &  Error\\
\hline
$\hbox{\PlikALL}$ &  $3.00$   & $_{-0.20}^{+0.19}$ & $_{-0.15}^{+0.16}$ &  $_{- 0.13}^{+0.12 }$ \\ 
$\hbox{\hlpALL}$ &  $2.87$    & $_{-0.14}^{+0.15}$ & $\pm 0.13$  &  $_{- 0.05 }^{+0.06 }$\\ 
$\hbox{\hlpALLps}$ &  $2.85$  & $\pm 0.14$ & $\pm 0.13$ & $_{- 0.05 }^{+0.05 }$\\ 
$\hbox{\hlpALLps(\Plik-like)}$ & $2.90$ & $_{-0.16}^{+0.17}$ &  $_{-0.15}^{+0.16}$  & $_{- 0.05 }^{+0.05 }$\\ 
\hline
\end{tabular}
\caption{Full error on \Neff\ and contributions from Statistics and Nuisance derived
using profile likelihoods and \CLASS\ (cf. description of the procedure in Section \ref{syst-statsec}) 
obtained when combining  \PlikALL, \hlpALL\ and \hlpALLps\ 
with \lowTEB\ and \BAO. \label{tab:syst-stat}}
\end{table}

From the comparison of the results of \hlpALLps(\Plik-like) and \hlpALL, we
can deduce that the additionnal data induce a slight shift ($\Delta\Neff=0.03$),
apart from the expected reduction of the statistical error.

From the comparison of the results of \hlpALLps(\Plik-like) and  \PlikALL,
we observe that the statistical error is exactly the same: giving high confidence to the fact
that the impact of the different choices made in the likelihood implementation for the covariance matrix 
is negligible. 
The remaining difference, which happens to be the bigger one, comes from the effect of the 
foreground modelling, which impacts  both
the mean value and the nuisance error. The foreground modelling (but a different one) is also tested
through the comparison
of the results of \hlpALL\ and \hlpALLps.




\subsection{Other cosmological data}

We have further tested the impact of CMB Lensing on the \Neff\ measurement and
found it to be very small, as expected \citep[cf.][]{planck2014-a15}, slightly
lowering the overall results by $0.01$. 

We have also checked that the
choice of the low-$\ell$ likelihood had no impact on the final results (replacing
\lowTEB\ with \lollipop+\Commander\ as stated in Sect.~\ref{sec:lowell}). 

It has to be noted that the supernovae data do not help 
to further constraint  \Neff\ once the \BAO\ data are  used, we therefore chose not to use them in this analysis.
For completeness we note that the update of the \BAO\ data from DR11 to DR12 does not
impact the constraint on \Neff\ \citep[][]{Alam:2016hwk}.

\subsection{Summary}
\begin{figure}[]
\centering
\includegraphics[width=6.5cm,angle=-90]{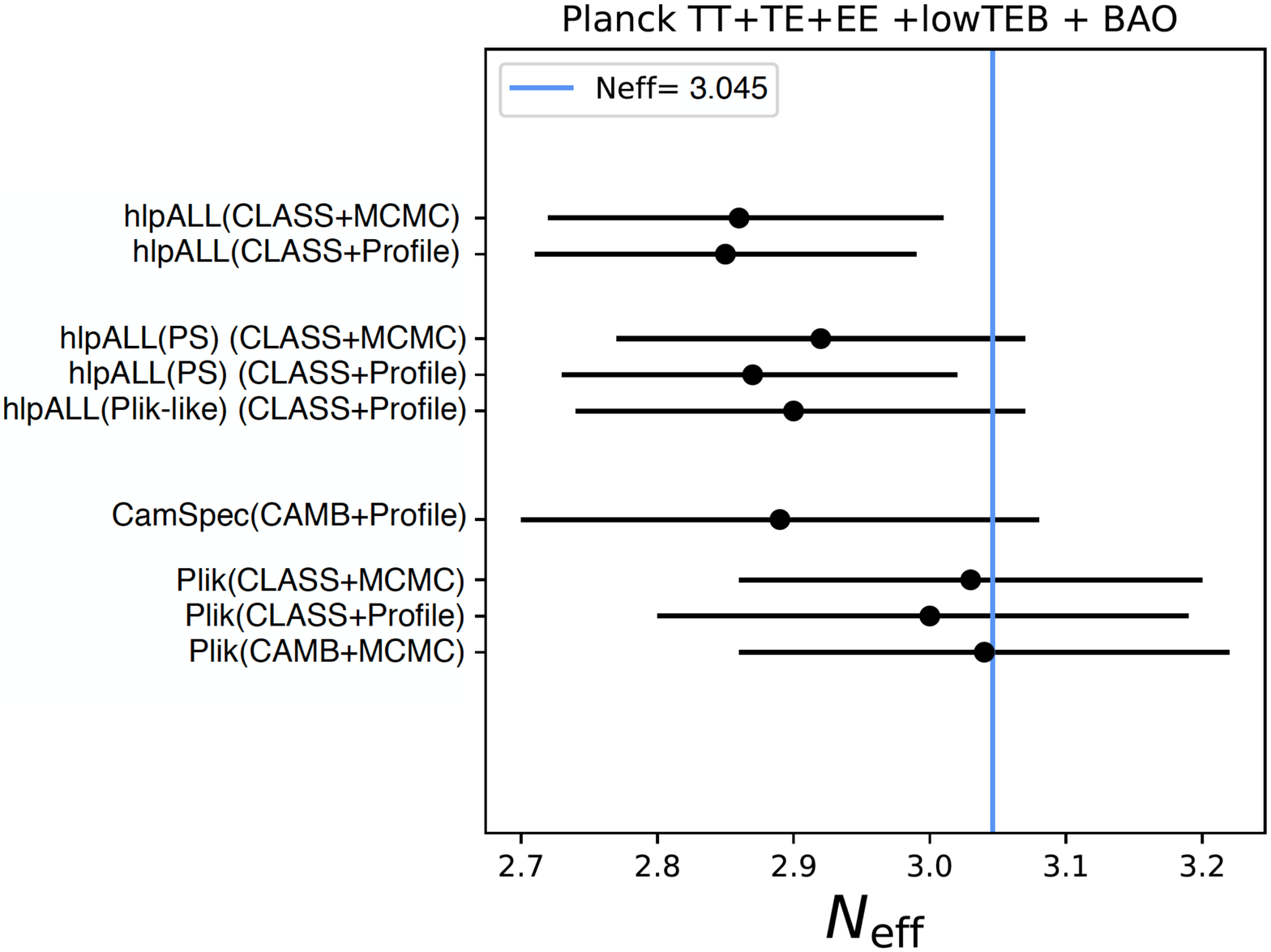}
\caption{Results comparisons for the combination of \planck\ $\rm{TT}$+$\rm{TE}$+$\rm{EE}$ likelihoods, 
with \lowTEB\ and \BAO.}
\label{fig:Neff_summary}
\end{figure}

The results on $\Neff$ are summarized in \myfigure~\ref{fig:Neff_summary}. 
The  shift of the mean values observed when using a likelihood or another is
of the same order of magnitude as the error derived from each individual likelihood
($\Delta\Neff\simeq 0.17$ vs. $\sigma(\Neff)=0.18$). 
It has been shown to be mainly driven by the assumptions made for the foreground modelling.
A small part of this variation (up to $0.03$) has been identified to be linked to
the data considered in \hlp\ and not in \plik. Still, it is high enough 
not to be neglected when constraining theoretical models from the \Neff\ measurement
only. 

\section{Fitting $\rm{TT}$ and $\rm{TE}$ separately}
\label{consistencysec}
In the previous sections we have shown the results of the combination of temperature
and polarisation CMB\ data. In the following, we estimate \Neff\ for 
$\rm{TT}$ and $\rm{TE}$ separately to further compare  the outcome
of each likelihood. 

\subsection{$\rm{TT}$+\lowTEB+\BAO\ results}

In this section, we consider the combination of temperature-only CMB likelihoods, together with \BAO\ data.
The results are summarized in \mytable~\ref{tab:TT_baosn} for various configurations. 
For this specific combination the \CamSpec\ results are not public, we therefore cannot use
them in the comparison.

\begin{table}
\begin{tabular}{l l c } 
\hline\hline       
\hbox{\planck} $\cal{L}$& Config & \Neff   \\
\hbox{+\lowTEB+\BAO}&  &  \\
\hline
$\hbox{\PlikTT\offi}$ &MCMC/\CAMB & $3.15\pm0.23$   \\ 
\hline
$\hbox{\PlikTT} $& Profile/\CLASS &   $3.09_{-0.22}^{+0.21}$   \\ 
$\hbox{\hlpTT }$& Profile/\CLASS &  $3.27_{-0.26}^{+0.28}$  \\ 
$\hbox{\hlpTTps }$& Profile/\CLASS & $3.20_{-0.20}^{+0.21}$ \\ 
\hline
\hline
\end{tabular}
\caption{Results on \Neff\ obtained when combining  \PlikTT, \hlpTT\ and \hlpTTps\ 
with \BAO\ (errors are given at 68$\%$CL). \lowTEB\ has been used at low-$\ell$.}  
\label{tab:TT_baosn} 
\end{table}

From this \mytable, we obtain $\Delta\Neff\simeq 0.18$
from the largest difference observed
between $\hbox{\hlpTT }$-Profile/\CLASS\ and
$\hbox{\PlikTT\offi}$-Profile/\CLASS. 

As in the previous section, we have checked that the
impact of the neutrino settings is almost negligible, as well as the impact
of supernovae data. In addition, the
choice of the DR12 \BAO\ data instead of DR11 has no effect.

\subsection{$\rm{TE}$+\lowTEB+\BAO\ results}
\label{SectionTE}
Given the \planck\ noise level, the $\rm{TE}$ likelihoods lead to similar results than those obtained with $\rm{TT}$ on \lambdacdm.
In addition they are less sensitive to the foreground modellings \citep[]{galli:2014,Couchot:2016vaq}. In this section we 
use of the $\rm{TE}$ likelihood in place of the $\rm{TT}$ one and compare the results
obtained on \Neff\ when combined with \lowTEB\ and \BAO\ on \mytable~\ref{tab:TE}. 

The remaining $\Delta\Neff$ is of the 
order of 0.07, which is small with respect to the total error with $\rm{TE}$ only. 
It may still contain some residual systematics from temperature to
polarisation leakage which study is beyond the scope of this paper.

\begin{table}
\begin{tabular}{l l c } 
\hline\hline       
\hbox{\planck} $\cal{L}$& Config & \lowTEB   \\
\hline
$\hbox{\PlikTE} $& MCMC/\CAMB  & $2.94\pm0.37$\\
\hline
$\hbox{\hlpTE} $& Profile/\CLASS & $3.01_{-0.30}^{+0.32}$ \\
\hline                  
\end{tabular}
\caption{Results on \Neff\ obtained when combining \BAO\ with \PlikTE, and \hlpTE\ (errors are given at 68$\%$CL).
\lowTEB\ has been used at low-$\ell$.}
\label{tab:TE}
\end{table}

\section{Conclusions}
\label{conclusec}

We have studied in detail the estimation of the effective number of relativistic
species from \CMB\ \planck\ data.
We have tested different ingredients of the analysis to further quantify their impact 
on the results: mainly the Boltzmann codes, the high$-\ell$ likelihoods (\plik, \hillipopps\ and
\camspec), and the statistical
analysis.  
\begin{itemize}
\item{} The estimated variation of \Neff\ when switching from \CAMB\ to \CLASS\ is 
negligible, of the order of $\Delta\Neff=0.01$. 
\item{} If we can safely neglect the impact of the covariance matrix estimation, as suggested by the
obtained results, the variation linked to the assumptions on foreground residuals modelling
derived from the comparison of the high$-\ell$ likelihoods has been estimated to be of the order of $\Delta\Neff=0.17$
on which a small part (up to $0.03$) may be attributed
to a statistical effect.  We have also shown that, at least for \hlp, $\Neff$ was mainly correlated 
with nuisance parameters linked to foregrounds playing
a role at small scales (ie. point sources and SZ). 
\item{} We have found slight differences between the Bayesian and the frequentist inferred mean values, linked to
particular likelihood volume effects. A shift between both methods has been estimated to be $\Delta\Neff\leq 0.05$.
\end{itemize}
As an overall conclusion, we have shown that the variation of the mean \Neff\ values
is non-negligible. This foreground related systematic uncertainty is of the same order of magnitude as the error derived for each 
individual likelihood.  In addition the results obtained with \hlp\ and
\camspec\ lead systematically to lower values than the ones derived from the public \planck\ likelihood.

We have cross-checked the consistency of the results when considering $\rm{TT}$ and $\rm{TE}$ separately.
When considering $\rm{TE}$ only (together with \BAO\ and \lowTEB), which is less sensitive to foreground residuals, 
this observed variation drops down to $\Delta\Neff=0.05$ for the likelihoods we have been able to compare.

We have shown that likelihood modelling is an important challenge for the current \planck\ measurements
for the \Neff\ interpretation, even for temperature data. The shift discussed in this paper
is very large compared to the $\sigma(\Neff)=0.027$ statistical-only expectations for 
CMB-S4. We however expect data from the next generation of CMB experiments to be more robust 
to such systematic error. The increase in constraining power from the $\rm{TE}$ power spectrum with respect to 
the $\rm{TT}$ one, as well as the better determination of the temperature power spectrum on 
small scales will reduce the impact of foregrounds mismodelling. 

\appendix{}
\section*{Appendix on nuisance parameters}
\setcounter{table}{0}
\renewcommand{\thetable}{A\arabic{table}}

This appendix presents the nuisance parameters of the \hillipopps\ likelihood in table \ref{tab:hlp_nuisance}.
\begin{table}[!ht]
\begin{center}
\begin{tabular}{ll}
\hline
\hline
name & definition \\
\hline
\multicolumn{2}{l}{\bf{Instrumental calibrations}}\\
\hline
$c^\planck_{100hm1}$ & map calibration (100-hm1)  \\  
$c^\planck_{100hm2}$ & map calibration (100-hm2)  \\  
$c^\planck_{143hm1}$ & map calibration (143-hm1)  \\  
$c^\planck_{143hm2}$ & map calibration (143-hm2)  \\  
$c^\planck_{217hm1}$ & map calibration (217-hm1)  \\  
$c^\planck_{217hm2}$ & map calibration (217-hm2)\\  
$A_\planck$ & absolute calibration  \\
\hline
\multicolumn{2}{c}{\bf{Foreground modellings}} \\
\hline
$A^\planck_{\rm PS}(100 \times 100)$	& PS amplitude in TT  (100x100 GHz) \\
$A^\planck_{\rm PS}(100 \times 143)$	& PS amplitude in TT (100x143 GHz) \\
$A^\planck_{\rm PS}(100 \times 217)$	& PS amplitude in TT (100x217 GHz) \\
$A^\planck_{\rm PS}(143 \times 143)$	& PS amplitude in TT (143x143 GHz) \\
$A^\planck_{\rm PS}(143 \times 217)$	& PS amplitude in TT (143x217 GHz) \\
$A^\planck_{\rm PS}(217 \times 217)$	& PS amplitude in TT (217x217 GHz) \\
$A_{\rm radio}$	& scaling  for radio sources (TT) \\
$A_{\rm dusty}$	& scaling  for infrared  sources (TT) \\
$A_{\rm SZ}$ & scaling for the tSZ template (TT) \\
$A_{\rm CIB}$ & scaling for the CIB template (TT)  \\
$A_{\rm kSZ}$ & scaling for the kSZ template (TT) \\
$A_{\rm SZxCIB}$ & scaling for kSZ x CIB cross correlation \\
$A_{\rm dust}^{\rm TT}$ & scaling  for the dust in TT \\
$A_{\rm dust}^{\rm EE}$ & scaling  for the dust in EE \\
$A_{\rm dust}^{\rm TE}$ & scaling  for the dust in TE \\
\hline
\hline
\end{tabular}
\caption{Nuisance parameters for the \hillipopps\ likelihood.}
\label{tab:hlp_nuisance}
\end{center}
\end{table}

\newpage
\bibliographystyle{aat}
\bibliography{refs2}

\end{document}

%% file: macros.tex
\def\CMB{{CMB}}

\def\myfigure{Fig.}
\def\mytable{Table}

\newcommand{\Plik}{{\tt Plik}}
\newcommand{\offi}{{$^{\sharp}$}}
\newcommand{\offismall}{{${\sharp}$}}
\newcommand{\xpol}{{\tt Xpol}}

\newcommand{\PlikTT}{{\tt PlikTT}}
\newcommand{\PlikTE}{{\tt PlikTE}}
\newcommand{\plikALL}{{\tt PlikALL}}
\newcommand{\PlikALL}{{\tt PlikALL}}

\newcommand{\hlp}{{\tt{HiLLiPOP(PS)}}}
\newcommand{\hillipopps}{{\tt{HiLLiPOP}}}
\newcommand{\hillipop}{{\hlp}}
\newcommand{\CAMEL}{\textsc{CAMEL}}
\newcommand{\hlpTT}{{\tt hlpTT(PS)}}
\newcommand{\hlpTTps}{{\tt hlpTT}}
\newcommand{\hlpXX}{{\tt hlp(PS)}}
\newcommand{\hlpXXps}{{\tt hlp}}
\newcommand{\hlpALL}{{\tt hlpALL(PS)}}
\newcommand{\hlpALLps}{{\tt hlpALL}}
\newcommand{\hlpTE}{{\tt hlpTE}}
\newcommand{\hlpPS}{{\tt hlpTT}}

\newcommand{\lollipop}{{\tt{Lollipop}}}
\newcommand{\Commander}{{\tt{Commander}}}

\newcommand{\mnu}{\ensuremath{\mathrm{\Sigma m_{\nu}}}}

\newcommand{\lambdaCDM}{\ensuremath{\mathrm{\Lambda{CDM}}}}

\newcommand{\Neff}{\ensuremath{N_\mathrm{{eff}}}}
\newcommand{\lambdacdm}{\lambdaCDM}

\newcommand{\p}{\ensuremath{\mathrm{p}}}

\newcommand{\CamSpec}{{\tt CamSpec}}

\newcommand{\CamSpecALL}{{\tt CamSpecALL}}
\newcommand{\camspec}{{\tt CamSpec}}
\newcommand{\plik}{{\tt Plik}}

\newcommand{\mksym}[1]{\ifmmode {\rm #1}\else #1\fi}

\newcommand{\BAO}{\mksym{BAO}}

\newcommand{\lowTEB}{\mksym{lowTEB}}

\newcommand{\taureio}{\ensuremath{\tau_{\rm reio}}}
\newcommand{\As}{\ensuremath{\mathrm{A_{\rm s}}}}

\newcommand{\Alens}{\ensuremath{A_{\rm L}}}

\providecommand{\planck}{\Planck}

\providecommand{\text}[1]{\rm{#1}}

\providecommand{\muK}{\mu\rm{K}}

\newcommand{\eV}{\,\text{eV}}

\providecommand{\CAMB}{{\tt CAMB}}

\providecommand{\CLASS}{{\tt CLASS}}

\newcommand{\begm}{\begin{pmatrix}}
\newcommand{\enm}{\end{pmatrix}}

\newcommand\ba{\begin{eqnarray}}
\newcommand\ea{\end{eqnarray}}
\newcommand\bea{\begin{eqnarray}}
\newcommand\eea{\end{eqnarray}}

\newcommand\be{\begin{equation}}
\newcommand\ee{\end{equation}}

\def\pmb#1{\setbox0=\hbox{#1}%
    \kern-.025em\copy0\kern-\wd0
    \kern.05em\copy0\kern-\wd0
    \kern-.025em\raise.0433em\box0}

\def\p2Y{\;_2Y}
\def\m2Y{\;_{-2}Y}
\def\beglet{
  \addtocounter{equation}{1}%
  \setcounter{parentequation}{\value{equation}}%
  \setcounter{equation}{0}%
  \def\theequation{\arabic{parentequation}\alph{equation}}%
  \ignorespaces
}
\def\endlet{
  \setcounter{equation}{\value{parentequation}}%
  \def\theequation{\arabic{equation}}%
}
\providecommand{\beglet}{\begin{subequations}}
\providecommand{\endlet}{\end{subequations}}

\def\setsymbol#1#2{\expandafter\def\csname #1\endcsname{#2}}
\def\getsymbol#1{\csname #1\endcsname}

\newbox\tablebox    \newdimen\tablewidth
\def\leaderfil{\leaders\hbox to 5pt{\hss.\hss}\hfil}
\def\tablenote#1 #2\par{\begingroup \parindent=0.8em
    \abovedisplayshortskip=0pt\belowdisplayshortskip=0pt
    \noindent
    $$\hss\vbox{\hsize\tablewidth \hangindent=\parindent \hangafter=1 \noindent
    \hbox to \parindent{$^#1$\hss}\strut#2\strut\par}\hss$$
    \endgroup}

\def\L2{\ifmmode L_2\else $L_2$\fi}

\def\DeltaT{\ifmmode \Delta T\else $\Delta T$\fi}
\def\deltat{\ifmmode \Delta t\else $\Delta t$\fi}
\def\fknee{\ifmmode f_{\rm knee}\else $f_{\rm knee}$\fi}
\def\Fmax{\ifmmode F_{\rm max}\else $F_{\rm max}$\fi}
\def\solar{\ifmmode{\rm M}_{\mathord\odot}\else${\rm M}_{\mathord\odot}$\fi}
\def\Msolar{\ifmmode{\rm M}_{\mathord\odot}\else${\rm M}_{\mathord\odot}$\fi}
\def\Lsolar{\ifmmode{\rm L}_{\mathord\odot}\else${\rm L}_{\mathord\odot}$\fi}
\def\inv{\ifmmode^{-1}\else$^{-1}$\fi}
\def\mo{\ifmmode^{-1}\else$^{-1}$\fi}
\def\sup#1{\ifmmode ^{\rm #1}\else $^{\rm #1}$\fi}
\def\expo#1{\ifmmode \times 10^{#1}\else $\times 10^{#1}$\fi}
\def\,{\thinspace}
\def\lsim{\mathrel{\raise .4ex\hbox{\rlap{$<$}\lower 1.2ex\hbox{$\sim$}}}}
\def\gsim{\mathrel{\raise .4ex\hbox{\rlap{$>$}\lower 1.2ex\hbox{$\sim$}}}}

\def\simprop{\mathrel{\raise .4ex\hbox{\rlap{$\propto$}\lower 1.2ex\hbox{$\sim$}}}}
\def\deg{\ifmmode^\circ\else$^\circ$\fi}
\def\pdeg{\ifmmode $\setbox0=\hbox{$^{\circ}$}\rlap{\hskip.11\wd0 .}$^{\circ}
          \else \setbox0=\hbox{$^{\circ}$}\rlap{\hskip.11\wd0 .}$^{\circ}$\fi}
\def\arcs{\ifmmode {^{\scriptstyle\prime\prime}}
          \else $^{\scriptstyle\prime\prime}$\fi}
\def\arcm{\ifmmode {^{\scriptstyle\prime}}
          \else $^{\scriptstyle\prime}$\fi}
\newdimen\sa  \newdimen\sb
\def\parcs{\sa=.07em \sb=.03em
     \ifmmode \hbox{\rlap{.}}^{\scriptstyle\prime\kern -\sb\prime}\hbox{\kern -\sa}
     \else \rlap{.}$^{\scriptstyle\prime\kern -\sb\prime}$\kern -\sa\fi}
\def\parcm{\sa=.08em \sb=.03em
     \ifmmode \hbox{\rlap{.}\kern\sa}^{\scriptstyle\prime}\hbox{\kern-\sb}
     \else \rlap{.}\kern\sa$^{\scriptstyle\prime}$\kern-\sb\fi}
\def\ra[#1 #2 #3.#4]{#1\sup{h}#2\sup{m}#3\sup{s}\llap.#4}
\def\dec[#1 #2 #3.#4]{#1\deg#2\arcm#3\arcs\llap.#4}
\def\deco[#1 #2 #3]{#1\deg#2\arcm#3\arcs}
\def\rra[#1 #2]{#1\sup{h}#2\sup{m}}

\def\dots{\relax\ifmmode \ldots\else $\ldots$\fi}
\def\WHzsr{\ifmmode $W\,Hz\mo\,sr\mo$\else W\,Hz\mo\,sr\mo\fi}
\def\mHz{\ifmmode $\,mHz$\else \,mHz\fi}
\def\GHz{\ifmmode $\,GHz$\else \,GHz\fi}
\def\mKs{\ifmmode $\,mK\,s$^{1/2}\else \,mK\,s$^{1/2}$\fi}
\def\muKs{\ifmmode \,\mu$K\,s$^{1/2}\else \,$\mu$K\,s$^{1/2}$\fi}
\def\muKRJs{\ifmmode \,\mu$K$_{\rm RJ}$\,s$^{1/2}\else \,$\mu$K$_{\rm RJ}$\,s$^{1/2}$\fi}
\def\muKHz{\ifmmode \,\mu$K\,Hz$^{-1/2}\else \,$\mu$K\,Hz$^{-1/2}$\fi}
\def\MJysr{\ifmmode \,$MJy\,sr\mo$\else \,MJy\,sr\mo\fi}
\def\MJysrmK{\ifmmode \,$MJy\,sr\mo$\,mK$_{\rm CMB}\mo\else \,MJy\,sr\mo\,mK$_{\rm CMB}\mo$\fi}
\def\microns{\ifmmode \,\mu$m$\else \,$\mu$m\fi}

\def\muK{\ifmmode \,\mu$K$\else \,$\mu$\hbox{K}\fi}
\def\microK{\ifmmode \,\mu$K$\else \,$\mu$\hbox{K}\fi}
\def\muW{\ifmmode \,\mu$W$\else \,$\mu$\hbox{W}\fi}
\def\kms{\ifmmode $\,km\,s$^{-1}\else \,km\,s$^{-1}$\fi}
\def\kmsMpc{\ifmmode $\,\kms\,Mpc\mo$\else \,\kms\,Mpc\mo\fi}
\providecommand{\sorthelp}[1]{}